# Ontologies for increasing the FAIRness of plant research data

Kathryn Dumschott[1], Hannah Dörpholz[1], Marie-Angélique Laporte[2], Dominik Brilhaus[3], Andrea Schrader[4], Björn Usadel[1,5], Steffen Neumann[6,7], Elizabeth Arnaud[2], and Angela Kranz[1]

**Affiliations**

[1] Institute of Bio- and Geosciences (IBG-4: Bioinformatics) & Bioeconomy Science Center (BioSC), CEPLAS, Forschungszentrum Jülich GmbH, 52425 Jülich, Germany

[2] Digital Inclusion, Bioversity International, 34397 Montpellier, France

[3] Data Science and Management & Cluster of Excellence on Plant Sciences (CEPLAS), Heinrich Heine University Düsseldorf, 40225 Düsseldorf, Germany

[4] Data Science and Management & Cluster of Excellence on Plant Sciences (CEPLAS), University of Cologne, 50674 Cologne, Germany

[5] Institute for Biological Data Science & Cluster of Excellence on Plant Sciences (CEPLAS), Heinrich Heine University Düsseldorf, 40225 Düsseldorf, Germany

[6] Leibniz Institute of Plant Biochemistry, 06120 Halle, Germany

[7] German Centre for Integrative Biodiversity Research (iDiv), 04103 Halle-Jena-Leipzig, Germany

**Key words**

Data management, metadata, FAIR, ontologies, OBO Foundry, DataPLANT, ISA

## Abstract

The importance of improving the FAIRness (findability, accessibility, interoperability, reusability) of research data is undeniable, especially in the face of large, complex datasets currently being produced by omics technologies. Facilitating the integration of a dataset with other types of data increases the likelihood of reuse, and the potential of answering novel research questions. Ontologies are a useful tool for semantically tagging datasets as adding relevant metadata increases the understanding of how data was produced and increases its interoperability. Ontologies provide concepts for a particular domain as well as the relationships between concepts. By tagging data with ontology terms, data becomes both human and machine interpretable, allowing for increased reuse and interoperability. However, the task of identifying ontologies relevant to a particular research domain or technology is challenging, especially within the diverse realm of fundamental plant research. In this review, we outline the ontologies most relevant to the fundamental plant sciences and how they can be used to annotate data related to plant-specific experiments within metadata frameworks, such as Investigation-Study-Assay (ISA). We also outline repositories and platforms most useful for identifying applicable ontologies or finding ontology terms.

# 1. Introduction

The field of plant research encompasses a huge range of experimental designs and analytical techniques in order to elucidate the complex, interconnected mechanisms involved in plant systems in a controlled manner, providing insights into the respective mechanisms and facilitating the development of new technologies and strategies for improving crop productivity, disease resistance, and environmental sustainability (Shah et al., 2019; Senger et al., 2022; Baekelandt et al., 2023).

Documenting the experimental designs and resulting research data ranges from capturing the experimental design to its implementation, from sample characteristics to experimental or environmental factors, and from capturing phenotyping and imaging data to molecular analyses such as genomics, transcriptomics, proteomics and metabolomics data. The size and complexity of such experimental designs and the resulting data challenge good data management practices and the preservation of data. Conversely, some investigations result in scarce amounts of data that may be significant if combined with additional datasets, so long as they are properly preserved. For this reason, the FAIR (findable, accessible, interoperable and reusable) principles were designed to guide data producers to maximize good data management practices (Wilkinson et al., 2016). FAIR data ensures transparency, reproducibility, and interoperability of plant science research, facilitating collaboration among scientists and enhancing the overall quality and impact of research outcomes. This in turn allows scientists to more easily contribute to, and more rapidly adapt to, the development of sustainable solutions for addressing global challenges such as food security and climate change (Mayer et al., 2021; Arend et al., 2022).

One key component of research data management (RDM) is the comprehensive and accurate description with metadata, or data about data. Metadata provides essential information about the context, content, and characteristics of the data, helping researchers to organize, describe, and understand datasets and their production, enabling effective data discovery, sharing, and reuse (Wilkinson et al., 2016). The correct and complete recording of metadata relating to an investigation is especially important for plant research data as environmental conditions can have such a profound influence over the resulting data of sessile organisms (Ćwiek-Kupczyńska et al., 2019).

With the large amounts of data being generated for a single research project, the potential and benefits for reusing and combining datasets to facilitate novel scientific discoveries is becoming ever greater. The challenge lies in the ability to find and integrate relevant datasets from different sources. Metadata is crucial for the correct interpretation of experimental data. Consistency in metadata annotation is important to ensure data is interoperable. It is crucial that datasets are both standardized as well as not only human- but also machine-readable, especially in the plant sciences where diverse types of data are collected, analyzed and integrated (Shaw et al., 2020; Pommier et al., 2023).

In recent years, ontologies have emerged as a powerful building block, supporting the standardization and harmonization of data annotation in plant sciences and increasing their FAIRness. Ontologies are systematic descriptions of knowledge used to describe a specific domain (Jensen and Bork, 2010). They are composed of a collection of terms as well as the relationships between them, which adds context and structure. Ontologies provide unique identifiers for concepts, making them machine-readable and retrievable. The standardized definitions for terms ensure that metadata tagged with ontology terms is interoperable between researchers.

In addition to ontologies, metadata frameworks are important and widely-used data models for increasing the interoperability and shareability of data (Sansone et al., 2012). Metadata frameworks promote the structuring of data, ensuring it is in a consistent format which allows both data producers and consumers to effectively work with a diverse dataset. Well known examples of metadata frameworks include lightweight Bioschemas (Michel and The Bioschemas Community, 2018) and the ISA metadata framework (Investigation-Study-Assay) (Sansone et al., 2012; González-Beltrán et al., 2014). Implementing metadata frameworks in conjunction with ontologies further facilitates the FAIRness of data.

In recent years, several large-scale efforts aiming to provide services and tools that contribute to increasing the FAIRness of research data have been organized. Examples of such efforts are the German National Research Data Infrastructure (NFDI) (https://www.nfdi.de/; Hartl et al., 2021) and Elixir (https://elixir-europe.org/; Crosswell and Thornton, 2012). The NFDI comprises 26 consortia from different scientific disciplines, many of which offer software that facilitate standardized and comprehensive metadata annotation (Sasse et al., 2022). For example, the plant-focused NFDI DataPLANT offers a metadata annotation tool chain, Swate, that incorporates ontologies necessary for the annotation of plant-specific experiments (Mühlhaus et al., 2021).

While there is no 'one size fits all' approach to correctly annotating data with ontology terms, the task of selecting a specific term or ontology is often daunting and confusing to newcomers due to diverse and scattered resources. To aid researchers in this task, this review will provide an overview of ontologies most relevant to the fundamental plant sciences as well as their role and application in the annotation and integration of plant-specific experiments and how they relate to metadata frameworks, such as ISA. We will outline repositories and platforms most helpful for finding ontologies or ontology terms applicable to the annotation of metadata in the fundamental plant sciences. Finally, we will discuss the importance of community engagement for the interoperability of ontologies and ensuring that ontologies reflect the most recent scientific advancements.

## 2. Ontologies for increased interoperability of research data

Ontologies are formal descriptions of knowledge that define concepts or terms and categories within a specific domain, as well as the relationships between them (Figure 1) (Gruber, 1993). While structure and semantic language of an ontology facilitates automatic reasoning, ontologies created within the biological sciences often focus on hierarchically describing their concepts and terms, meaning that they can be used as a well-organized controlled vocabulary for a specific domain. In the context of RDM, they are important for the structuring and standardization of data, improving interoperability and facilitating data integration and reuse.

Ontologies need to fulfill certain technical requirements and structures to contribute to standardized metadata annotation. A few technical terms that need to be understood by users can be found below and in Figure 1A:

- **Classes** are a primary component of ontologies. They define general concepts, terms or types of objects within a particular scientific domain. For example, terms such as 'plant structure' and 'abiotic plant exposure' are concepts that could be necessary to annotate a plant-related experiment.

- **Individuals**, also known as instances or particulars, are specific examples of a class (Hoehndorf et al., 2015). For example, "*Arabidopsis thaliana*", a plant species, is an individual of the class "species".
- **Properties**, also known as relations, describe how ontology classes and individuals are connected. They create semantic context within ontologies, giving them (often hierarchical) structure. The most commonly found relation in an ontology is SubClassOf (is_a). The descriptiveness of these relations or properties allows tools to reason or infer information from an ontology.

Classes contain information, or annotations, about the particular concept or term, including a human-readable name or label, definition, equivalent terms, and synonyms. Most importantly, ontology classes contain a unique identifier, such as a persistent identifier (PID) or a Uniform Resource Identifier (URI). These are permanent and unique, which allows the term to be permanently identified (Osumi-Sutherland, 2020). An ontology term's PID can be used to 'tag' metadata within a document, thereby making it machine-readable (Figure 1B).

Properties are used to create semantic context within ontologies, and can be subdivided into three groups: object properties, data properties and annotation properties. Object properties describe the relationships between classes or individuals (Figure 1A), thereby giving an ontology its overall structure (as shown in Figure 1C). Some of the most commonly used object properties are SubClass Of (is_a) and part_of. Data properties describe the relationships between a class and the type of data value an associated variable has, also known as a datatype. For example, the class 'publication' would have the datatype 'string' (a sequence of characters) for the 'author name' and the datatype 'DATE' (a calendar date in a format such as "YYYY-MM-DD" for the year, month and day respectively) for 'publication date'. Annotation properties link a class with the annotations of the class (described above) and provide the PID. Unlike object and data properties, annotation properties are ignored by semantic reasoners.

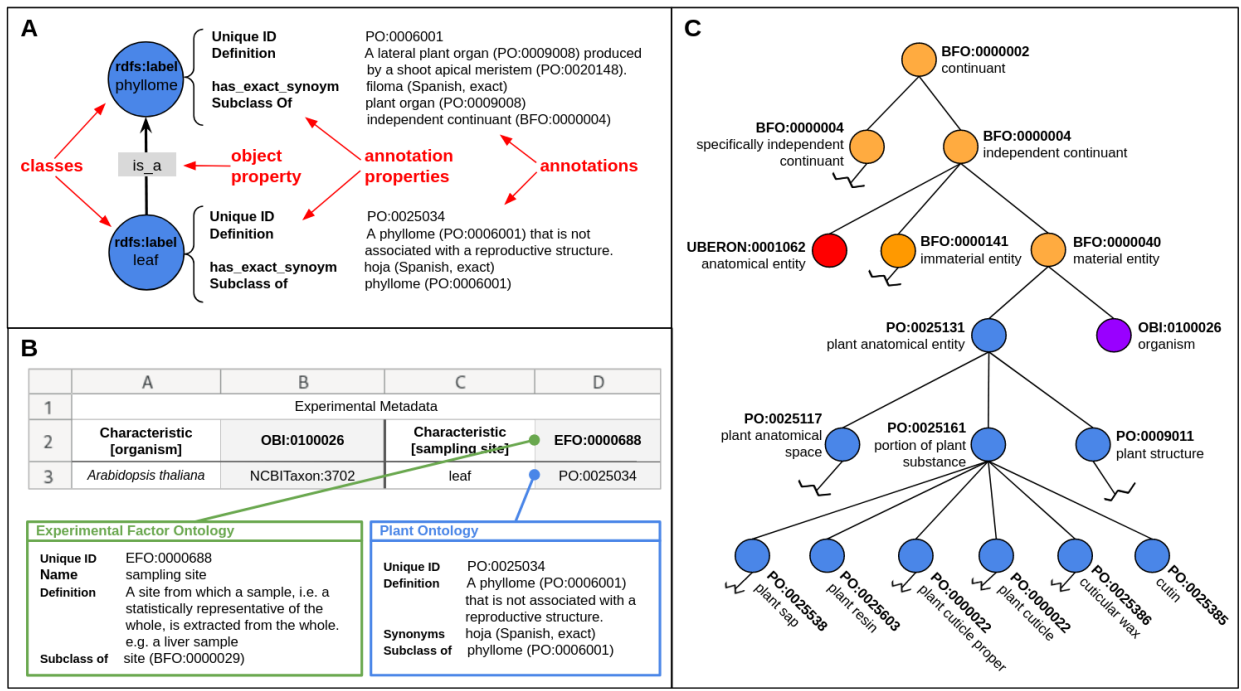

**Figure 1:** Ontology structure and functions. (A) Ontology classes (blue circles) contain information such as a unique identifier, name, definition and synonyms for describing the class. Classes are connected via properties, specifically object properties (arrow labeled "is_a") which give structure and context. (B) An example of how ontology terms can be incorporated into metadata sheets to 'tag' information, facilitating machine-readability and increasing the FAIRness of the data. Columns A and C depict terms and columns B and D depict the corresponding ontology ID, respectively. Row 2 depicts the experiment characteristic and row 3 is the corresponding value (C) Ontologies can import and share terms from other ontologies to enable consistent representation of a term (concept) or domain, increasing the interoperability and standardization of ontologies and the terms they contain. Pictured here is an excerpt of the Plant Ontology (PO) (Walls et al., 2012; Cooper et al., 2013), containing terms from the Basic Formal Ontology (BFO) (Arp et al., 2015), the Uberon multi-species anatomy ontology (UBERON) (Mungall et al., 2012) and the Ontology for Biomedical Investigations (OBI) (Brinkman et al., 2010).

Ontologies can be expressed in a number of ways which vary in their human-readability and usage. The most commonly expressed forms within the biological sciences are the Open Biological and Biomedical Ontology (OBO) file format and the Web Ontology Language (OWL) (Antoniou et al., 2009; Golbreich et al., 2007). Originally developed by the World Wide Web Consortium (W3C; https://www.w3.org/), OWL is a semantic language that can be used by many tools. There are number of different dialects of OWL, including OWL/XML (Hori et al., 2003), OWL Functional Syntax (Glimm et al., 2009) and OWL Manchester Syntax (Horridge et al., 2006), which vary in human-readability. The OBO file format was originally designed for ontologies within the biomedical sciences. Concepts modeled within OBO file format are loosely based off of OWL and the format aims at human-readability and minimal redundancy

([https://owlcollab.github.io/oboformat/doc/GO.format.obo-1_2.html](https://owlcollab.github.io/oboformat/doc/GO.format.obo-1_2.html)). However, converting from OWL to OBO format often results in errors and must be done with caution.

While ontologies vary in terms of content, one nearly ubiquitous characteristic of ontologies, especially within the biomedical and plant sciences, is the open and collaborative mindset with which its developers work to increase interoperability. One way of accomplishing this is to reuse, or import, terms from an existing ontology if the term adequately describes what is needed (Xiang et al., 2010), or else to use import from an upper-level ontology to organize terms at the most general level (Figure 1C). Upper-level ontologies provide a foundational framework for other ontologies to incorporate in order to create semantic interoperability (Hoehndorf, 2010,). As they are independent of any particular domain or application and provide the most general concepts, so that they can be easily integrated into ontologies and facilitate data integration across different systems and domains (Mascardi et al., 2007). For example, a 'plant anatomical entity' (PO:0025131) can be sorted under the upper-level ontology term 'material entity' (BFO:0000040) (Figure 1C). Other reasons for importing terms are, for example, commonly used measurement units, or the creation of a very specific application ontology, which may require terms from a related domain ontology to give proper context or structure.

An additional benefit of term reuse and importing is that even if an ontology project runs out of support or funding, proper integration and cross-referencing of terms in other ontologies ensures that their knowledge is not lost. Interoperability can also be increased via community engagement. Contributions from domain experts (researchers) help grow and improve ontologies, keeping them up to date as the science itself develops and scientific discoveries are made. These days, many ontologies are available on GitHub ([https://github.com/](https://github.com/)) in order to facilitate open source work and versioning control.

## 3. Ontology Resources

While the benefit of tagging metadata with ontology terms is clear, it is often less clear determining how to select an appropriate ontology or term. Over recent years, a number of resources involved with the collection, curation and development of ontologies for particular scientific domains have been developed. These are often good starting points when deciding what term is best for metadata annotation. As they are easily searchable and often give the current developmental status of the ontologies they include, researchers can find the information they need to select a suitable ontology, or ontology term, for their particular data.

Among these is the Open Biological and Biomedical Ontology (OBO) Foundry (https://obofoundry.org/; Smith et al. 2007), a collaborative initiative with the aim of developing a set of interoperable ontologies intended for the biological sciences. The OBO Foundry is a development community with a set of guiding principles that are seen as good practice, working to increase interoperability of ontologies. The principles are improved and refined at regular intervals and many are operational, allowing for easier interpretation and adherence (Jackson et al., 2021). The principles cover all aspects of ontology development, from licensing and format to documentation and commitment to collaboration. Ontologies wishing to be accepted into the OBO community are checked against these principles before being accepted. While not a classic ontology repository, the ontologies included in the OBO registry are considered to be adhering to best ontology practices and that terms and relations found within are actively maintained. The registry covers a wide range of ontologies spanning general topics such as biological processes, molecular entities and scientific protocols and investigations, including a number of plant-focused ontologies.

While the OBO Foundry includes a number of plant-related ontologies, topics more relating to the plant sciences, such as plant genomics, phenomics, agronomy related domains are not in the focus and lack the same level of community involvement as the biomedical domain ontologies. For this reason, a number of ontology repositories specifically geared to different facets of the plant sciences have been developed. Among these is AgroPortal, a vocabulary and ontology repository founded by the Montpellier scientific community to facilitate open and collaborative science in agronomy (http://agroportal.lirmm.fr/; Jonquet et al., 2018). AgroPortal was specifically designed for agronomy and related domains and reuses the openly available Ontoportal software (https://ontoportal.org/; Jonquet et al. 2023) to build the repository and services platform. The project aims to provide a reliable service involving hosting, searching and improving ontologies, allowing users to actively participate in the platform by uploading content and commenting on others' content (Jonquet et al., 2018). The original motivation for AgroPortal was guided by five agronomic use cases, which cover a range of agronomic topics from germplasm to livestock and contribute to the design and focus of the portal (https://agroportal.lirmm.fr/about). These use cases include the Agronomic Linked Data knowledge base (http://agrold.southgreen.fr/aldp/; Venkatesan et al., 2018), a knowledge-based database for plant molecular networks such as genes, proteins, metabolic pathways and plant traits and the Crop Ontology (CO) Project (https://cropontology.org/; Arnaud et al., 2022) of the Integrated Breeding Platform (https://www.integratedbreeding.net/), described in greater detail below. AgroPortal includes projects, vocabularies and ontologies which cover the entire range of agronomic research, from livestock and plant species to environmental conditions and land governance.

The Planteome database is heavily based on ontologies and an informative resource for scientists searching for terminologies applicable to plant research and describing plant traits and experiments (https://planteome.org/). The database contains a collection of general reference ontologies aimed at improving annotation of an array of plant related research data, ranging from genes to phenotypes (Cooper et al., 2018). Planteome also actively maps the species-specific ontologies of the Crop Ontology against the species-neutral reference ontologies, allowing users to search for a trait without having to consider the specific species. This is particularly useful for studies in comparative genomics or investigations of a family or clade (Cooper et al., 2018). As with the previously mentioned platforms, the Planteome ontologies are publicly available and openly maintained via GitHub repository (https://github.com/Planteome) to encourage sharing, tracking of revisions and new term requests (Cooper et al., 2018).

A number of additional repositories for finding and querying ontologies are also available and include Ontobee (https://ontobee.org/; Ong et al., 2011), BioPortal (https://www.bioontology.org/; Whetzel et al., 2011), The Ontology Lookup Service (https://www.ebi.ac.uk/ols/index; Côté et al., 2006), and the newly developed TIB Terminology service (https://terminology.tib.eu/ts; Anders et al., 2022). While their collections do not focus heavily on the plant sciences, all aim to facilitate data sharing, ontology visualization, querying, integration and analysis. As the plant sciences cover a wide range of different technologies, many of the ontologies collected in these repositories will contain terms relevant to experimental set up and analysis. For example, all repositories listed above contain the Chemical Methods Ontology (CHMO) (https://github.com/rsc-ontologies/rsc-cmo), an ontology developed to describe chemical methods applicable to experimental assays, such as electron microscopy, preparations of materials to be separated for further analysis, such as by electrophoresis and the synthesis of materials. Terms included in CHMO are relevant to many common techniques used throughout plant-related experiments, making it crucial for metadata annotation for plant researchers.

# 4. The landscape of ontologies for fundamental plant science

One challenge of identifying ontologies relevant to fundamental plant research is the diverse range of methods and technologies that can and are utilized for an average investigation or research project. The mechanisms governing processes such as plant development or resistance to stress and disease are complex and oftentimes a variety of techniques spanning different scientific fields are required to comprehensively elucidate these pathways and responses. For example, physiological changes such as photosynthesis, chemical changes to the soluble leaf fraction, and changes in gene expression of target pathways must all be combined to comprehensively characterize how progressive soil drought stress influences sugar alcohol accumulation in soybean (Dumschott et al., 2019). To properly annotate the metadata accompanying such a study, ontology domains covering plant traits, experimental conditions, experimental protocols, equipment and technologies, and measurement units must all be incorporated (Figure 2). In this section, we will outline the ontologies most relevant to the fundamental plant sciences, divided the section into two subsections: general scientific ontologies and plant-related ontologies. A list of relevant ontologies and their domains can be found in Table 1. An extended version is available in Supplementary Table S1.

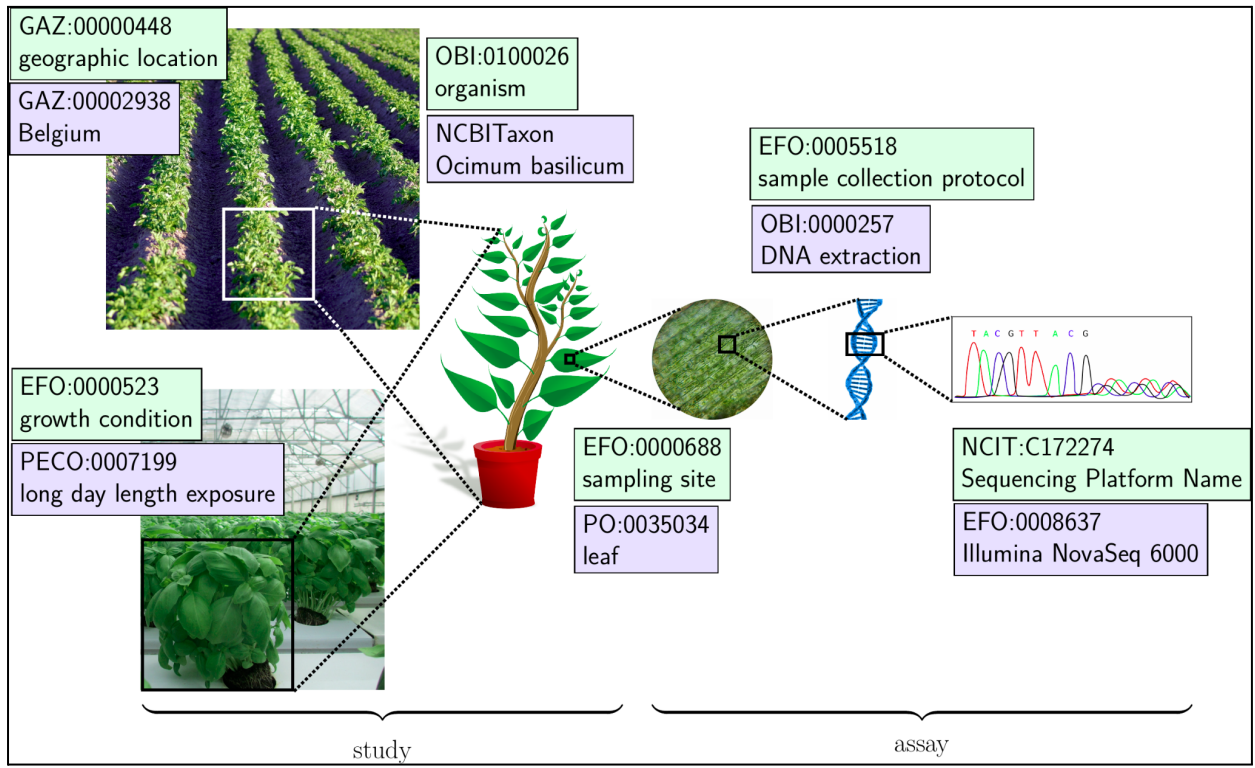

**Figure 2:** Example of how ontologies can be used for annotating a plant science experiment to increase the FAIRness of data. Terms must adequately cover the location and condition under which the plants are grown, the samples collected, and how the samples were processed and analyzed. For this reason, a diversity of ontologies is required to capture the multidimensional nature of the experiment. Also shown are the ISA metadata sections describing the growth and sample preparation protocols and samples in the *Study* and the protocols and research data related to an *Assay*.

## 4.1. General scientific ontologies

General scientific ontologies cover concepts, relationships, and properties within a particular domain that can be applicable to wider scientific fields. We identified both upper-level and domain-focused ontologies as being relevant and important for the annotation of research data within the fundamental plant sciences. These ontologies provide a common vocabulary for representing and integrating scientific knowledge from different scientific domains, covering topics such as experimental set-ups or measurement units used, or else provide a framework for unifying ontologies for increased interoperability. In this section, we outline examples of general scientific ontologies, both domain specific and upper-level ontologies, that are most relevant to contemporary fundamental plant research.

For upper-level ontologies, we identified BFO, COB, and the Relation Ontology (RO) as being the most relevant to the fundamental plant sciences. BFO is an upper-level ontology that provides a framework for building more specific domain ontologies. It aims to provide a common foundation of concepts and relationships that can be used to represent knowledge in a wide range of domains such as biology, biomedical informatics, natural language processing, and philosophy (Arp et al., 2015). COB is a basic and structured framework that serves as a foundational resource for the life sciences (https://github.com/OBOFoundry/COB). By capturing essential concepts and their relationships, it serves as a foundation for the development of advanced ontologies for biology and biomedicine (Abdelmageed et al., 2021). A general scientific ontology that defines all properties used by the OBO Foundry ontologies is the RO. It is a formal ontology that provides a framework for interoperability between different ontologies and can therefore be used for different contexts (Smith et al,. 2005).

One crucial aspect of successfully understanding the context of a dataset, is understanding how an experiment was performed. Variations in what treatment was performed on a sample, what protocols were used to extract a target material, or how a sample was collected must all be properly annotated for the resulting data and analysis to have any meaning. For this reason, the Ontology for Biomedical Investigation (OBI) (Brinkman et al., 2010) and the Experimental Factor Ontology (EFO) (Malone et al., 2010) were developed. OBI is an integrated ontology for the description of all aspects of life-science investigations, covering all phases of investigations, from planning to reporting (Bandrowski et al., 2016). OBI reuses terms from other well-established ontologies, in order to cover the wide range of projects it is intended for. It can be used to annotate metadata for investigations involving omics and multi-omics approaches, and incorporates information and material entities that represent processes, roles and functions of information. OBI is a member ontology of the OBO Foundry, encouraging cross-discipline collaboration and standardization. EFO is developed and maintained by the European Bioinformatics Institute (EMBL-EBI) to support the annotation, analysis and visualization of data handled by groups involved in the organization, focusing on gene expression data (Malone et al., 2010).

When annotating investigations involving molecular pathways and the characterization of gene expression and function, ontologies such as the Gene Ontology (GO) (Ashburner et al., 2000) and MapMan (Schwacke et al. 2019) are good resources to consider. GO is a widely used standardized classification system that provides a controlled vocabulary and framework for describing the functions, processes, and cellular components associated with genes and gene products, such as proteins, across different species (Ashburner et al., 2000; Gene Ontology Consortium, 2023). It helps researchers annotate and interpret genomic data by assigning functional terms to genes based on experimental evidence and computational predictions.

Terms within the ontology are divided into three main categories: biological processes, molecular function and cellular component. Terms are connected within these categories, forming directed acyclic graphs. GO enables the comparison of gene function across diverse organisms, facilitates data integration and analysis, and supports the discovery of new biological insights by providing a structured and comprehensive framework for studying gene functions in the context of biological systems.

In contrast to GO, the MapMan4 ontology was developed specifically for the characterization of gene expression and biological functions in plants (Schwacke et al., 2019). Built upon the original MapMan framework (Thimm et al., 2004), MapMan4 represents common biological processes and genetic information gathered from a wide range of plant species. It is organized in a tree structure, with top levels being main biological concepts and subsequent sublevels becoming more specialized to ensure the most precise protein characterization possible. The tool Mercator is used for the automatic annotation of protein sequences with the MapMan4 ontology (Schwacke et al., 2019).

As the field of fundamental plant research incorporates many different scientific domains, there are a number of ontologies in other natural and life-science disciplines that are often applicable for a plant science investigation, depending on the topic and scope. Chemical Entities of Biological Interest (ChEBI) (Degtyarenko et al., 2007) and the Environment Ontology (ENVO) (Buttigieg et al., 2013, 2016), covering chemical and environmental aspects, respectively, are two such often-utilized, cross-discipline ontologies. Developed by EMBL-EBI, ChEBI provides a classification of chemical entities (Degtyarenko et al., 2008). To increase its interoperability with ontologies in the biomedical domain, ChEBI provides a mapping (in the form of a bridge OWL) to the BFO (Hastings et al., 2013). The ontology can be divided into three main branches: chemical entities, the role the chemical entity can have, and subatomic particles. Its importance to the wider scientific community is evident as ChEBI is widely incorporated into various databases, such as UniProt (The UniProt Consortium, 2015), and is heavily reused in well-established ontologies, such as GO, OBI, and EFO as well as plant-specific ontologies. For environmental entities, ENVO is a widely-used ontology for describing ecosystems, environmental processes or even entire planets. Originally designed to provide information regarding biomes and environmental features of genomic and microbiome samples for the Genomics Standards Consortium (Field et al., 2011), ENVO has since evolved into a cross-discipline resource, spanning domains from biomedicine and omics to anthropogenic ecology and socioeconomic development (Buttigieg et al., 2016). As ENVO evolved, developers implemented changes to better align to OBO Foundry principles to ensure increased interoperability, such as incorporating terms from RO and BFO and moving the ontology to its own GitHub repository for better versioning control.

Finally, there are a number of ontologies specifically relating relevant metadata important for describing the measurement of samples and subsequent data analysis. For example, the Human Proteome Organization- Proteomics Standards Initiative (HUPO-PSI) developed the PSI-Mass Spectrometry (MS) controlled vocabulary to logically structure and capture all terms relating to an MS pipeline- from sample preparation to instrument parts, parameters and related software (Mayer et al., 2013). Examples of ontologies for the annotation of software used during the analysis of data collected within the life sciences are the Software Ontology (SWO) (Malone et al., 2014) and the Ontology of Bioscientific Data Analysis and Data Management (EDAM) (Ison et al., 2013). The need for such ontologies was realized as bioinformatic analysis became ever more prevalent. Just as metadata annotation is crucial for the reproducibility of laboratory experiments, knowing what versions of what tools were used to analyze a data set is necessary for the reproducibility of said analyses. The scope of SWO is broad as it incorporates tools and software versions not only used in bioinformatics analyses, but also tools (and their versions) used for the management, analysis and presentation of biological data (Malone et al., 2014). EDAM covers topics, operations, types of data and data identifiers and formats relevant to data analysis and management in the life sciences (Ison et al., 2013).

While a number of ontologies and their applications have been described above and in Table 1, they are just a small subset of ontologies available to fundamental plant scientists. Depending on the particular investigation and technologies being employed, researchers may require an ontology that covers a domain not covered in this review, like food and nutritional ontologies. For this reason, it is always recommended to consult established ontology repositories to find a term or ontology that most closely matches the metadata being annotated.

## 4.2. Plant-specific ontologies

There are several ontologies well suited for describing and annotating experiments, phenotypic traits, structures and experimental conditions relating to plant research. In the following section, we will describe the most relevant ones in greater detail.

*The Planteome Project reference ontologies*

The Planteome project (https://planteome.org/; Cooper et al., 2018) develops and maintains a number of species-neutral reference ontologies, including the Plant Ontology (PO) (Walls et al., 2012; Cooper et al., 2013), the Plant Trait Ontology (TO) (Jaiswal, 2011) and the Plant Experimental Conditions Ontology (PECO) (Cooper et al., 2018). The PO is crucial for the consistent annotation of anatomy, morphology and developmental stages of both plants and plant parts (Walls et al., 2012). Originally focused on *Arabidopsis thaliana* (mouse-ear cress) *Zea mays* (corn) and *Oryza sativa* (rice), it was broadened to cover all Viridiplantae (green plants). The primary aim of the PO is to bridge the diversity of data that can be collected during plant research- from genetics, molecular and cellular biology to taxonomy and botany research (Walls et al., 2019). The PO is divided into two main branches: 'plant anatomical entity' and 'plant structure development stage'. Terms in these branches are organized hierarchically via subclass, or subClassOf (is_a) relations. All other relations depicted in the PO come from the OBO RO (Walls et al., 2019). The branch 'plant anatomical entity' includes terms for plant morphology and anatomy, such as structures (Walls et al., 2012), whereas the branch 'plant structure development stage' covers terms relating to stages of life either of a whole plant or plant part during which the structure undergoes developmental processes (i.e. growth, differentiation or senescence) (Walls et al., 2019).

Two other plant reference ontologies used in the Planteome database are the TO and PECO. Both ontologies were conceived and developed with the aim of improving data interoperability for the advancement of plant research. One common issue within the fundamental plant sciences is the semantic inconsistencies that exist between species, especially for phenotypic descriptions, meaning data integration is often not possible without the manual identification of corresponding concepts (Arnaud et al., 2012). For example, what is referred to as a 'leaf' in some species is known as a 'frond' in others. It is therefore crucial to standardize trait terms between different species and projects in a way that allows for the easy integration of phenotypic and trait data from different sources. The TO was developed to address the discrepancies in trait descriptions and to increase interoperability of plant trait data between species (Cooper et al., 2018). Terms within the TO are structured according to an Entity-Quality pattern (Arnaud et al., 2012). Entity terms are imported from other well-established ontologies, such as the PO, the GO and ChEBI, while quality terms are taken from the Phenotype and Trait Ontology (PATO). In this way, terms and descriptions are kept general enough that they can be successfully applied to most plant species and importing terms from other ontologies facilitates its interoperability. Finally, PECO covers terms specifically needed to describe the growth conditions used in a variety of plant experiments, including abiotic or biotic treatments assessed during an experiment (Cooper et al., 2018).

*The Crop Ontology (CO)*

The Crop Ontology (CO) was developed by several members of the Consultative Group on International Agricultural Research (CGIAR) to harmonize the annotation of phenotypic and genotypic data between different crops (https://alliancebioversityciat.org/tools-innovations/crop-ontology). As traits, measurement methods and scales can vary greatly between different crops, controlled vocabularies and ontologies are required to enable comparisons between how a trait is assessed in different species (Shrestha et al., 2012). The CO provides crop-specific trait ontologies for increased plant data annotation and integration (Shrestha et al., 2012; Arnaud et al., 2016; Arnaud et al., 2020). At the time of writing, 37 species-specific ontologies are included, covering a wide range of different crop species, including staple crops (wheat, maize, cotton, soybean), fruits and vegetables (banana, brassica) and legumes (chickpea, mung bean, lentil, faba bean).

*Plant Phenology Ontology (PPO)*

While a number of large continental-scale data sources for plant phenology exist, the ability to conduct analysis of plant phenology on an inter-continental scale was hindered by the lack of standardized language and terminology used to describe the data found in individual repositories, resulting in data incompatibility. For this reason, the Plant Phenology Ontology (PPO) was developed to address this communication gap and help to facilitate interoperability of plant phenology data (Stucky et al., 2018). Six principles guide the design and goals of the PPO to ensure it is both broadly applicable as well as interoperable and that data based on PPO annotation is reusable. The PPO aims to reuse terms from other ontologies, such as the PO and Biological Collections Ontology (BCO) wherever possible (Walls et al., 2014). The classes and concepts included in the PPO can be divided into three main topics: plant structures, phenological traits and observations of/data about phenological traits (Stucky et al., 2018).

*The Plant Phenotype Experiment Ontology (PPEO)*

One of the most challenging fields of fundamental plant research is that of phenotyping, due to its heterogeneous nature and the sensitivity of phenotype to environmental conditions. The ability to correctly interpret phenotypic data is therefore heavily reliant on how completely environmental conditions and metadata relating to experiment setups is recorded (Papoutsoglou et al., 2023). Therefore, Ćwiek-Kupczyńska et al. 2016 created the "Minimum Information About a Plant Phenotyping Experiment" (MIAPPE) to outline the list of attributes, or metadata, necessary to adequately annotate a plant phenotyping experiment so that the resulting data can be correctly understood. Attributes included within MIAPPE are organized into different sections according to the ISA framework (described in more detail in Section 5) as it is able to handle a diverse range of phenotyping data and experimental designs due to its generality and flexibility (Ćwiek-Kupczyńska et al., 2016). To facilitate the implementation of MIAPPE, the Plant Phenotyping Experiment Ontology (PPEO) was created. Whereas most of the plant-specific ontologies described above focus on defining concepts and building relations around them, PPEO focuses first on the structure of MIAPPE, incorporating the different sections of the framework as the primary backbone of the ontology, then adding the required attributes (https://github.com/MIAPPE/MIAPPE-ontology).

**Table 1:** Commonly used ontologies for plant science research. While in most cases, the ontology term ID space is the same across different portals, in a few cases e.g. OBO Foundry and BioPortal use different short names. An extended version is available in Supplementary Table S1.

| Ontology ID | Ontology Name | Domain | References |
|---|---|---|---|
| **Plant-specific ontologies** | | | |
| AGRO | AGRonomy Ontology | Agronomic practices, techniques, and variables used in agronomic experiments. | Devare et al. 2016 Aubert, 2017 |
| CO | Crop Ontology | Breeder's traits and variables | Shrestha et al., 2012; Arnaud et al., 2016; Arnaud et al., 2022 |
| FLOPO | Flora Phenotype Ontology | Traits and phenotypes of flowering plants | Hoehndorf et al. 2016 |
| PECO | Plant Experimental Conditions Ontology | Plant experimental conditions | Cooper et al. 2018 |
| PO | Plant Ontology | Plant anatomy, morphology and growth and development | Walls et al. 2012;Walls et al. 2019;Cooper et al. 2013 |
| PPEO | Plant Phenotype Experiment Ontology | Plant Phenotypes and Traits (implementation of the Minimal | Papoutsoglou et al. 2020 |

|  |  | Information About Plant Phenotyping Experiment) |  |
|---|---|---|---|
| PPO | Plant Phenology Ontology | Phenology of individual plants and populations of plants | Stucky et al. 2018 |
| PSO (OBO Foundry and OLS) | Plant Stress Ontology | Biotic and abiotic stresses that a plant may encounter | Cooper et al. 2018 |
| TO | Plant Trait Ontology | Phenotypic traits in plants | Jaiswal 2011 |
|  |  |  |  |
| **General scientific ontologies – Domain-specific** | | | |
| BAO | BioAssay Ontology | Biological screening assays and their results | Abeyruwan et al., 2014 |
| BCO | Biological Collections Ontology | Support the interoperability of biodiversity data, including data on museum collections, environmental/metagenomic samples, and ecological surveys | Walls et al. 2014 |
| BOF | Biodiversity Ontology | Biodiversity, as developed by the National Institute for Amazonian Research | Albuquerque 2011 |
| BTO | BRENDA Tissue and Enzyme Source Ontology | Source of an enzyme comprising tissues, cell lines, cell types and cell cultures | Gremse et al., 2011 |
| CHEBI | Chemical Entities of Biological Interest Ontology | Molecular entities of biological interest | Degtyarenko et al. 2008 |
| CHMO | Chemical Methods Ontology | Methods used to collect data in chemical experiments | https://github.com/rsc-ontologies/rsc-cmo |
| EDAM | EDAM Ontology of Bioscientific Data Analysis and Data Management | Computational biology, bioinformatics and bioimage informatics | Ison et al., 2013 |
| EFO | Experimental Factor Ontology | Experimental variables | Malone et al. 2010 |
| ENVO | Environment Ontology | Environmental systems, components and processes | Buttigieg et al. 2013, 2016 |
| GEOSPECIES | GeoSpecies Ontology | Integration of species concepts with species occurrences, gene sequences, images, references and geographical information | http://lod.geospecies.org/ |
| GO | Gene Ontology | Function of genes and gene products | Ashburner et al. 2000;Gene Ontology Consortium 2023 |

| MMO | Measurement Method Ontology | Methods used to make clinical and phenotype measurements | Smith et al., 2013 |
|---|---|---|---|
| MOD | Protein modification (PSI-MOD) | Protein chemical modifications, classified by molecular structure or amino acid | Montecchi-Palazzi et al. 2008 |
| MS | PSI Mass Spectrometry Ontology | proteomics mass spectrometry | Mayer et al., 2013 |
| MSIO | Metabolomics Standards Initiative Ontology (MSIO) | mass-spectrometry and nmr-spectroscopy based metabolomics experiments and fluxomics studies | Rocca-Serra, 2018 |
| NCBITAXON | National Center for Biotechnology Information (NCBI) Organismal Classification | NCBI organismal taxonomy | Federhen 2012 |
| NCIT | National Cancer Institute Thesaurus | Broad coverage of the cancer domain | https://ncit.nci.nih.gov |
| OBI | Ontology for Biomedical Investigations | Life-science and clinical investigations | Brinkman et al. 2010; Bandrowski et al. 2016 |
| PATO | Phenotype And Trait Ontology | Phenotypic qualities (properties, attributes or characteristics) | Gkoutos et al. 2018 |
| PCO | Population and Community Ontology | Material entities, qualities, and processes related to collections of interacting organisms such as populations and communities | Walls et al. 2014 |
| STATO | Statistics Ontology | statistical tests, conditions of application, and information needed or resulting from statistical methods | https://github.com/ISA-tools/stato |
| SWO | Software Ontology | software tools, their types, tasks, versions, provenance and associated data | Malone et al., 2014 |
| UBERON | Uberon multi-species anatomy ontology | An integrated cross-species anatomy, covers animals and bridges multiple species-specific ontologies | Mungall et al., 2012 |
| UO | Unit Ontology | Metrical units for use in conjunction with PATO | Gkoutos et al. 2012 |
| **General scientific ontologies – Upper-level ontologies** | | | |
| BFO | Basic Formal Ontology | Standardizes upper-level structure of OBO ontologies | Arp et al. 2015 |

| COB | Core Ontology for Biology and Biomedicine | Framework for building ontologies in the life sciences | https://github.com/OBOFoundry/COB |
| RO | Relation Ontology | Standardizes relations in OBO Foundry ontologies | https://oborel.github.io/obo-relations/ |

# 5. Implementation of metadata frameworks for ontology-enriched metadata annotation

Challenges in data harmony and integration often arise when bringing together data from multiple sources to answer complex scientific questions. These challenges involve differences in terminology descriptions, a lack of sufficient context, or else the data is structured in a way being difficult to comprehend, thereby hindering its reuse. To address these challenges and facilitate the management and integration of experimental data across different research domains, a number of community-driven efforts have been founded that aim to develop metadata standards and frameworks for improved data sharing and handling. Some well-known examples are Bioschemas (Michel F and The Bioschemas Community, 2018) and the ISA (Investigation-Study-Assay) framework (González-Beltrán et al., 2014, Sansone et al., 2016).

The ISA framework aims to ensure scientific data is accompanied by metadata that describes the context in which the data was generated. This context includes information about the experimental design, sample characteristics, data acquisition and processing, and data analysis. The ISA data model is structured around three entities: Investigation, Study, and Assay (Figure 3). The *Investigation* entity represents the overarching research project and provides general information about the research questions and goals. The *Study* entity describes the experimental design and methods used to collect the data, including information about the samples, treatments, and measurements. The *Assay* entity represents the specific data generated for each study, including raw and derived data files. One advantage of the ISA framework is its flexibility and adaptability, which allows for a wide range of experimental designs and research data to be represented within a single investigation. Its flexibility is exemplified by the ability to gather metadata in a user-friendly spreadsheet format, which is not only human-readable but also effortlessly convertible into machine-readable formats essential for numerous applications and software systems (e.g. ISA-Tab, ISA-JSON etc.) (González-Beltrán et al., 2014). Different types of biological data, such as transcriptomic, proteomic, or metabolomic data, as well as non-biological data, such as environmental conditions, can be represented by the ISA model (Sansone et al., 2012).

Ontologies play a crucial role in the ISA data model by providing a standardized vocabulary and sets of concepts that can be used to describe experimental metadata in a consistent and structured way. Each ISA entity has a list of required metadata, where ontology annotations are preferred over free-text to give the proper context for the entity (Figure 3A). According to the ISA specification, each entity has additional nodes to describe material or data related to the experiment, where ontology terms can further be annotated. For example, under the *Assay* entity is the 'material node', where researchers can describe materials consumed or produced during an experimental workflow. Within this node is the property "material type", which should be described using an ontology annotation (Figure 3B). This makes the description of experimental metadata both precise and unambiguous, reducing the potential for errors and misinterpretation, enabling data to be more easily integrated and shared across different research projects, domains and platforms. Ontologies can also be used to map the relationships between different concepts and terms, which can facilitate potential connections and correlations between different data sets.

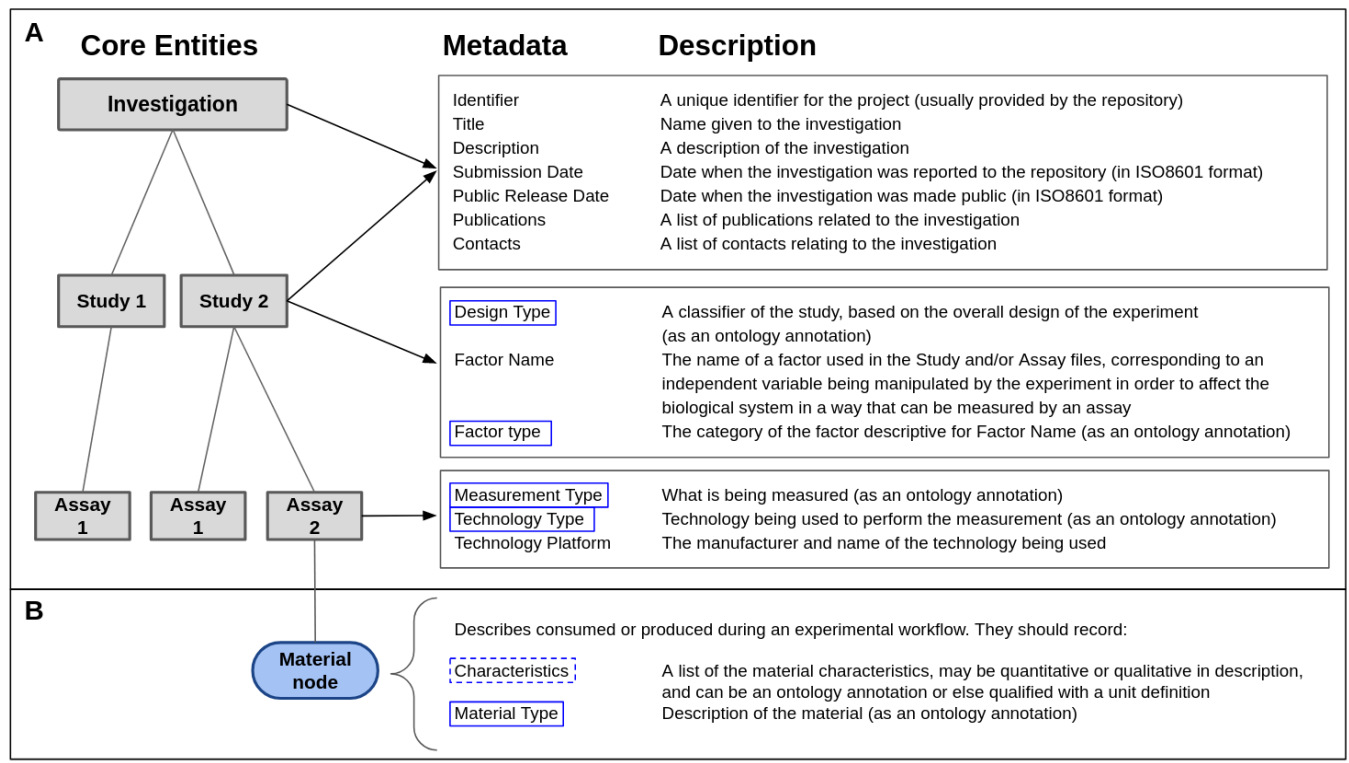

**Figure 3:** The ISA metadata framework (Sansone et al., 2016), designed to capture experimental metadata. (A) The framework consists of three core entities: *Investigation*, *Study* and *Assay*. The structure allows for multiple studies to be described within one investigation and multiple assays to be described within one study. Each entity has metadata requirements that should be included to ensure the complete description of the entity. Within the *Study* and *Assay* entities, an ontology annotation is the required input for included metadata, such as design type in study and measurement type in assay (marked with blue boxes). (B) An example of the 'material node' that *Assay* entities can contain to describe material consumed or produced during an experiment. The required metadata for this node are 'characteristics,' which may-but does not have to be- an ontology annotation (marked with a dashed blue box) and 'material type,' which required an ontology annotation (marked with a blue box).

There are a number of ways that plant-related ontologies can be incorporated into one of the ISA entities depending on the investigation that is being represented (Figure 4). For example, OBI and EFO can be incorporated into the *Study* and *Assay* entities of the ISA framework as they contain terms relating to experimental setup, sample processing and analysis. ENVO is particularly relevant for annotating *Study* metadata as terms can be used when describing the habitat or environmental conditions of a location when a plant sample was collected (for example, ENVO_1000745: drought).

One of the key benefits of using ontologies in the ISA data model is that it allows for more effective data integration and analysis, by controlling the values that a metadata element can take (for example Ho Sui et al., 2013; Peters et al., 2018). Using ontologies to standardize the terms used to describe experimental variables allows researchers to more easily compare data sets and identify similarities and differences between them. This can be especially important when analyzing large data sets, where manual inspection and interpretation of the data may be difficult or time-consuming. In addition to providing a standardized vocabulary, ontologies also help to improve the accuracy and consistency of the data itself. By using an ontology to specify the units of measurement used in an experiment, researchers can avoid errors that might arise from using different units or from converting units incorrectly. This can be especially important when working with complex data sets that involve many different types of measurements and units.

Overall, ontologies are an important tool for enhancing the effectiveness and efficiency of the ISA data model (Johnson et al., 2021). By providing a standardized vocabulary and set of concepts, ontologies help to ensure that the experimental metadata is precise, consistent, and easy to interpret and share across different research projects and domains.

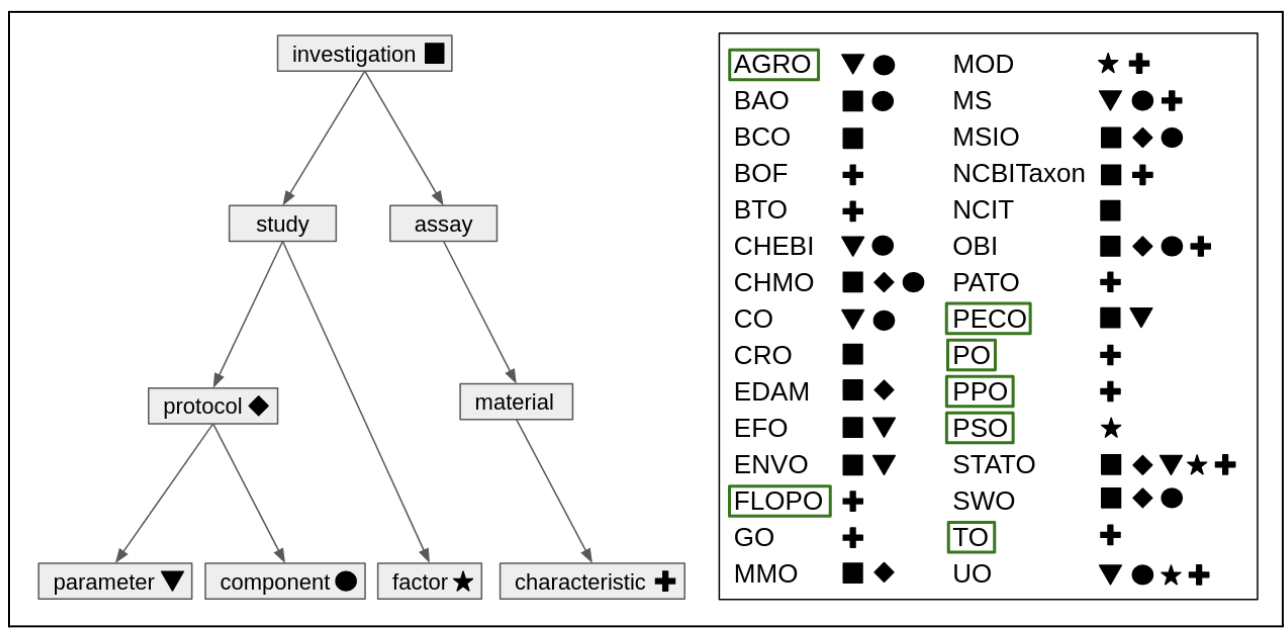

**Figure 4:** Visualization of the domain-specific ontologies relevant for the plant science community and their relation to ISA. Left: hierarchical structure of ISA concepts. The ontologies on the right are linked to ISA concepts via Symbols. Green boxes highlight plant-specific ontologies. While this figure gives an idea of where ontologies can be incorporated within ISA, there are likely scenarios where an ontology can be incorporated that is not depicted here.

# 6. Discussion and perspectives

As the potential of integrating modern techniques becomes evident, the need to properly integrate and manage the data produced becomes evermore important. Ontologies play a crucial role in this management, ensuring that data is both reusable and interoperable by 'tagging' data. Tagged data is then both human- and machine-readable, allowing for the subsequent retrieval and standardization of data. Incorporating terms into metadata frameworks, such as ISA, further increases the FAIRness of data However, despite the obvious advantages of incorporating ontologies into data management schemes and the fact that there is no 'one size fits all approach' to what terms should be used to annotate metadata, determining what ontology or ontological term is most appropriate at any given time can be challenging.

Fortunately, there has been a push in recent years to provide more comprehensive overviews of ontologies that are available to the wider scientific community. A number of ontology repositories have been developed, all aiming to facilitate  Ontology repositories and service platforms such as Planteome and AgroPortal are good resources for users trying to determine where to begin when annotating fundamental plant research data with ontology terms. Both are designed specifically for plant research, agronomy and related fields and encourage the active participation and collaboration of users for the improvement of included ontologies. Researchers can utilize the search function to find the best term to annotate data with.

One common theme apparent throughout the ontologies covered within this review is the importance of open and collaborative efforts for increasing the FAIRness of data. Ontology providers rely on and encourage community engagement and interdisciplinary collaboration for the continuing expansion and improvement of concepts and their relations. Open licensing and standardized ontology languages facilitate interoperability of ontologies and communication between different domain experts, which in turn increases reusability of ontologies within wider scientific communities. Fundamental plant research is diverse by nature, incorporating techniques and concepts from all corners of the life sciences, and benefits greatly from this community push for the standardization and improvement of ontologies for metadata annotation. The NFDI funded DataPLANT consortium recognizes the need of community engagement for the advancement of ontologies and closing the gap between novel technologies and terms used and proper semantic integration. For this reason, DataPLANT has developed its own ontology, called the DataPLANT Biology Ontology (DPBO) (https://github.com/nfdi4plants/nfdi4plants_ontology), which aims to close the ontology gap by providing quick feedback to term suggestions, and incorporating them into the DPBO. Ontology curators at DataPLANT then work to suggest the terms to well-known, established ontologies to be incorporated. In this way, researchers can use their needed terms and DataPLANT can act as the middle man between researchers and ontology providers. With the recent push for sustainable RDM and FAIR data practices, the potential that ontologies provide becomes evermore apparent. Proper annotation of metadata with ontology terms will no doubt further these practices, not only for the fundamental plant research community, but the wider scientific research community as a whole.

# Funding

We acknowledge support for DataPLANT 442077441 and NFDI4Chem 441958208 through the German National Research Data Initiative and CEPLAS is supported by Deutsche Forschungsgemeinschaft within the Excellence Initiative (EXC 1028) and under Germany's Excellence Strategy – EXC 2048/1 – project 390686111.

Jonquet, C., Toulet, A., Arnaud, E., Aubin, S., Dzalé Yeumo, E., Emonet, V., et al. (2018). AgroPortal: A vocabulary and ontology repository for agronomy. *Comput. Electron. Agric.* 144, 126-143. doi: 10.1016/j.compag.2017.10.012.

Jonquet, C., Graybeal, J., Bouazzouni, S., Dorf, M., Fiore, N., Kechagioglou, X., et al. (2023). Ontology Repositories and Semantic Artefact Catalogues with the OntoPortal Technology. 22nd International Semantic Web Conference (ISWC). Athens, Greece, France.

Malone, J., Brown, A., Lister, A. L., Ison, J., Hull, D., Parkinson, H., et al. (2014). The Software Ontology (SWO): a resource for reproducibility in biomedical data analysis, curation and digital preservation. *J. Biomed. Semant.* 5 (1), 1-13. doi: 10.1186/2041-1480-5-25.

Malone, J., Holloway, E., Adamusiak, T., Kapushesky, M., Zheng, J., Kolesnikov, N., et al. (2010). Modeling sample variables with an Experimental Factor Ontology. *Bioinformatics* 26 (8), 1112-1118. doi: 10.1093/bioinformatics/btq099.

Mascardi, V., Cordì, V. and Rosso, P. (2007). A Comparison of Upper Ontologies. In Workshop From Objects to Agents.

Mayer, G., Montecchi-Palazzi, L., Ovelleiro, D., Jones, A. R., Binz, P.-A., Deutsch, E. W., et al. (2013). The HUPO proteomics standards initiative-mass spectrometry controlled vocabulary. *Database* 2013, bat009. doi: 10.1093/database/bat009.

Mayer, G., Müller, W., Schork, K., Uszkoreit, J., Weidemann, A., Wittig, U., et al. (2021). Implementing FAIR data management within the German Network for Bioinformatics Infrastructure (de.NBI) exemplified by selected use cases. *Brief. Bioinform.* 22 (5). doi: 10.1093/bib/bbab010.

Michel, F. and The Bioschemas Community (2018). Bioschemas & Schema.org: a Lightweight Semantic Layer for Life Sciences Websites. *Biodivers. Inf. Sci. Stand.* 2, e25836. doi: 10.3897/biss.2.25836.

Montecchi-Palazzi, L., Beavis, R., Binz, P. A., Chalkley, R. J., Cottrell, J., Creasy, D., et al. (2008). The PSI-MOD community standard for representation of protein modification data. *Nat. Biotechnol.* 26 (8), 864-866. doi: 10.1038/nbt0808-864.

| Ontology ID | Ontology Name | Domain | License | Modularity | Obofoundry URL | OLS URL | BioPortal URL | Other URL |
|---|---|---|---|---|---|---|---|---|
| **Plant-Specific Ontologies** | | | | | | | | |
| AGRO | AGRonomy Ontology | Agronomic practices, techniques, and variables used in agronomic experiments. | CC BY 4.0 | OWL | https://obofoundry.org/ontology/agro.html | https://www.ebi.ac.uk/ols/ontologies/agro | https://bioportal.bioontology.org/ontologies/AGRO | |
| CO | Crop Ontology | Breeder's traits and variables | CC BY 4.0 | OWL | | | | https://cropontology.org/ |
| FLOPO | Flora Phenotype Ontology | Traits and phenotypes of flowering plants | CC0 1.0 | OWL | https://obofoundry.org/ontology/flopo.html | https://www.ebi.ac.uk/ols/ontologies/flopo | https://bioportal.bioontology.org/ontologies/FLOPO | |
| PECO | Plant Experimental Conditions Ontology | Plant experimental conditions | CC BY 4.0 | OWL | https://obofoundry.org/ontology/peco.html | https://www.ebi.ac.uk/ols/ontologies/peco | https://bioportal.bioontology.org/ontologies/PECO | |
| PO | Plant Ontology | Plant anatomy, morphology and growth and development | CC BY 4.0 | OWL and OBO | https://obofoundry.org/ontology/po.html | http://www.ebi.ac.uk/ontology-lookup/browse.do?ontName=PO | https://bioportal.bioontology.org/ontologies/PO | |
| PPEO | Plant Phenotype Experiment Ontology | Plant Phenotypes and Traits (implementation of the Minimal Information About Plant Phenotyping Experiment) | CC-BY 4.0 | OWL | --- | --- | --- | https://agroportal.lirmm.fr/ontologies/PPEO , https://github.com/MIAPPE/MIAPPE-ontology , https://fairsharing.org/1234 |
| PPO | Plant Phenology Ontology | Phenology of individual plants and populations of plants | CC BY 3.0 | OWL | https://obofoundry.org/ontology/ppo.html | https://www.ebi.ac.uk/ols/ontologies/ppo | https://bioportal.bioontology.org/ontologies/PPO | |
| PSO (OBO Foundry and OLS) PLANTSO (Bioportal) | Plant Stress Ontology | Biotic and abiotic stresses that a plant may encounter | CC BY 3.0 | OWL and OBO | https://obofoundry.org/ontology/pso.html | https://www.ebi.ac.uk/ols/ontologies/pso | https://bioportal.bioontology.org/ontologies/PLANTSO | |
| TO (OBO Foundry and OLS) PTO (Bioportal) | Plant Trait Ontology | Phenotypic traits in plants | CC BY 4.0 | OWL and OBO | https://obofoundry.org/ontology/to.html | https://www.ebi.ac.uk/ols/ontologies/to | https://bioportal.bioontology.org/ontologies/PTO | --- |
| **General Scientific Ontologies- Domain-Specific** | | | | | | | | |
| BAO | BioAssay Ontology | Biological screening assays and their results | CC-BY-SA 4.0 | OWL | --- | https://www.ebi.ac.uk/ols/ontologies/bao | https://bioportal.bioontology.org/ontologies/BAO | |
| BCO | Biological Collections Ontology | Support the interoperability of biodiversity data, including data on museum collections, environmental/metagenomic samples, and ecological surveys | CC0 1.0 | OWL | https://obofoundry.org/ontology/bco.html | https://www.ebi.ac.uk/ols/ontologies/bco | https://bioportal.bioontology.org/ontologies/BCO | |
| BTO | BRENDA Tissue and Enzyme Source Ontology | Source of an enzyme comprising tissues, cell lines, cell types and cell cultures | CC BY 4.0 | OWL and OBO | https://obofoundry.org/ontology/bto.html | https://www.ebi.ac.uk/ols/ontologies/bto | https://bioportal.bioontology.org/ontologies/BTO | |
| CHEBI | Chemical Entities of Biological Interest Ontology | Molecular entities of biological interest | CC BY 4.0 | OWL and OBO | https://obofoundry.org/ontology/chebi.html | https://www.ebi.ac.uk/ols/ontologies/chebi | https://bioportal.bioontology.org/ontologies/CHEBI | |
| CHMO | Chemical Methods Ontology | Methods used to collect data in chemical experiments | CC BY 4.0 | OWL | https://obofoundry.org/ontology/chmo.html | https://www.ebi.ac.uk/ols/ontologies/chmo | https://bioportal.bioontology.org/ontologies/CHMO | |
| EDAM | EDAM bioinformatics operations, data types, formats, identifiers and topics | Computational biology, bioinformatics and bioimage informatics | CC-BY-SA 4.0 | OWL | --- | https://www.ebi.ac.uk/ols/ontologies/edam | https://bioportal.bioontology.org/ontologies/EDAM | |
| EFO | Experimental Factor Ontology | Experimental variables | Copyright [2014] EMBL-European Bioinformatics Institute Licensed under the Apache License, version 2.0 (the "License") | OWL | --- | https://www.ebi.ac.uk/ols/ontologies/efo | https://bioportal.bioontology.org/ontologies/EFO | |
| ENVO | Environment Ontology | Environmental systems, components and processes | CC0 1.0 | OWL | https://obofoundry.org/ontology/envo.html | https://www.ebi.ac.uk/ols/ontologies/envo | https://bioportal.bioontology.org/ontologies/ENVO | |
| GEOSPECIES | GeoSpecies Ontology | Integration of species concepts with species occurrences, gene sequences, images, references and geographical information | CC BY 4.0 | OWL | --- | | https://bioportal.bioontology.org/ontologies/GEOSPECIES | |
| GO | Gene Ontology | Function of genes and gene products | CC BY 4.0 | OWL and OBO | https://obofoundry.org/ontology/go.html | https://www.ebi.ac.uk/ols/ontologies/go | | |
| MMO | Measurement Method Ontology | Methods used to make clinical and phenotype measurements | CC0 1.0 | OWL and OBO | https://obofoundry.org/ontology/mmo.html | https://www.ebi.ac.uk/ols/ontologies/mmo | https://bioportal.bioontology.org/ontologies/MMO | |
| MOD | Protein modification (PSI-MOD) | Protein chemical modifications, classified by molecular structure or amino acid | CC BY 4.0 | OWL and OBO | https://obofoundry.org/ontology/mod.html | | | |
| MS | PSI Mass Spectrometry Ontology | proteomics mass spectrometry | CC BY 3.0 | OWL and OBO | https://obofoundry.org/ontology/ms.html | https://www.ebi.ac.uk/ols/ontologies/ms | https://bioportal.bioontology.org/ontologies/MS | |
| MSIO | Metabolomics Standards Initiative Ontology (MSIO) | mass-spectrometry and nmr-spectroscopy based metabolomics experiments and fluxomics studies | CC BY 3.0 | OWL | --- | https://www.ebi.ac.uk/ols/ontologies/msio | --- | |
| NCBITAXON | National Center for Biotechnology Information (NCBI) Organismal Classification | NCBI organismal taxonomy | CC0 1.0 | OWL and OBO | https://obofoundry.org/ontology/ncbitaxon.html | https://www.ebi.ac.uk/ols/ontologies/ncbitaxon | https://bioportal.bioontology.org/ontologies/NCBITAXON | |
| NCIT | National Cancer Institute Thesaurus | Broad coverage of the cancer domain | CC BY 4.0 | OWL and OBO | https://obofoundry.org/ontology/ncit.html | https://www.ebi.ac.uk/ols/ontologies/ncit | https://bioportal.bioontology.org/ontologies/NCIT | |
| OBI | Ontology for Biomedical Investigations | Life-science and clinical investigations | CC BY 4.0 | OWL and OBO | https://obofoundry.org/ontology/obi.html | https://www.ebi.ac.uk/ols/ontologies/obi | https://bioportal.bioontology.org/ontologies/OBI | |
| PATO | Phenotype And Trait Ontology | Phenotypic qualities (properties, attributes or characteristics) | CC BY 3.0 | OWL and OBO | https://obofoundry.org/ontology/pato.html | https://www.ebi.ac.uk/ols/ontologies/pato | https://bioportal.bioontology.org/ontologies/PATO | |
| PCO | Population and Community Ontology | Material entities, qualities, and processes related to collections of interacting organisms such as populations and communities | CC0 1.0 | OWL | https://obofoundry.org/ontology/pco.html | https://www.ebi.ac.uk/ols/ontologies/pco | https://bioportal.bioontology.org/ontologies/PCO | |
| STATO | Statistics Ontology | statistical tests, conditions of application, and information needed or resulting from statistical methods | CC BY 3.0 | OWL | https://obofoundry.org/ontology/stato.html | https://www.ebi.ac.uk/ols/ontologies/stato | https://bioportal.bioontology.org/ontologies/STATO | |
| SWO | Software Ontology | software tools, their types, tasks, versions, provenance and associated data | CC BY 4.0 | OWL | https://obofoundry.org/ontology/swo.html | https://www.ebi.ac.uk/ols/ontologies/swo | | |
| UBERON | Uberon multi-species anatomy ontology | An integrated cross-species anatomy, covers animals and bridges multiple species-specific ontologies | CC BY 3.0 | OWL and OBO | | | | |
| UO | Unit Ontology | Metrical units for use in conjunction with PATO | CC BY 3.0 | OWL and OBO | https://obofoundry.org/ontology/uo.html | https://www.ebi.ac.uk/ols/ontologies/uo | https://bioportal.bioontology.org/ontologies/UO | |
| **Upper level ontologies:** | | | | | | | | |
| BFO | Basic Formal Ontology | Standardizes upper-level structure of OBO ontologies | CC BY 4.0 | OWL and OBO | https://obofoundry.org/ontology/ro.html | | https://bioportal.bioontology.org/ontologies/BFO | |
| COB | Core Ontology for Biology and Biomedicine | Framework for building ontologies in the life sciences | CC0 1.0 | OWL | https://obofoundry.org/ontology/cob.html | | | |
| RO | Relations Ontology | Standardizes relations in OBO Foundry ontologies | CC0 1.0 | OWL and OBO | https://obofoundry.org/ontology/ro.html | https://www.ebi.ac.uk/ols/ontologies/ro | | |